\documentclass[11pt,twoside]{article}
\usepackage{asp2010}

\newcommand{\msol}{\ensuremath{\rm{M}_{\odot}}}
\newcommand{\teff}{\ensuremath{T_{\rm{eff}}}}
\newcommand{\logg}{\ensuremath{\log g}}
\newcommand{\logy}{\ensuremath{\log y}}
\newcommand{\lheh}{\ensuremath{\log N_{\mathrm{He}}/N_{\mathrm{H}}}}

\newcommand{\uHz}{\ensuremath{\mu{\rm{Hz}}}}
\newcommand{\kep}{{\em Kepler}}
\def\ellone{$\ell$\,=\,1}
\def\elltwo{$\ell$\,=\,2}
\def\ellthree{$\ell$\,=\,3}
\def\emmzero{$m$\,=\,0}
\def\emmone{$m$\,=\,1}
\def\emmtwo{$m$\,=\,2}
\defcitealias{ostensen12b}{{\O}12}

\resetcounters

\begin{document}

\title{A helium-rich g-mode pulsator on the blue horizontal branch}
\author{Roy~H.~{\O}stensen}
\affil{Instituut voor Sterrenkunde, KU~Leuven,
B-3001 Leuven, Belgium}

\begin{abstract}
We introduce the first g-mode pulsator found to reside on the classical blue
horizontal branch. We extend the original discovery dataset to include more
than 3 years of continuous photometry of KIC\,1718290, which reveals a rich
spectrum of low-amplitude modes with periods between one and twelve hours,
most of which follow a regular spacing of 276.3 s. This mode structure
strongly resembles that of the V1093 Her pulsators, with only a slight shift
towards longer periods. Our spectroscopy, however, reveals KIC 1718290 to be
quite distinct from the sdB stars that show V1093Her pulsations, which all
have surface gravities higher than log g = 5.1 and helium abundances depleted
by at least an order of magnitude relative to the solar composition. We find
that KIC 1718290 has \teff\,=\,22\,100\,K, \logg\,=\,4.72, and a supersolar helium
abundance (\lheh\,=\,--0.45). This places it well above the extreme
horizontal branch, and rather on the very blue end of the classical
horizontal branch. We conclude that KIC 1718290 must have suffered extreme
mass loss during its first giant stage, but not sufficient to reach the
extreme horizontal branch.
\end{abstract}

\section{Introduction}

The {\em Kepler} target KIC\,1718290 (hereafter dubbed Largo)
appears in the Sloan survey as
SDSS J192300.68+371504.4 with {\em ugriz} magnitudes 15.5, 15.3, 15.5, 15.7, 15.9.
With these colours it was not UV-bright enough to be flagged as an sdB star
according to the selection criteria of the survey for hot subdwarf stars in
the {\em Kepler} field \citep{ostensen10b,ostensen11b}. Indeed, when it was
later targeted as part of a small survey for WD pulsators \citep{ostensen11c}
the spectrum reveled it to be a relatively He-rich star on the hot end of
the blue horizontal branch \citep[][hereafter {\O}121]{ostensen12b}. 
While the star was not targeted in the original search for compact pulsators,
it was observed in long-cadence mode (30-minute sampling) as an exoplanet target,
ensuring near-continous photometric coverage from quarters Q1 through Q16. 

Largo resides in the more crowded part of the Kepler field, at a galactic latitude
of $b$\,=\,+10$^\circ$, making it likely that it originates from a population of
younger, more metal-rich stars than the extreme horizontal branch stars found
further from the galactic plane.

While the discovery paper was based on {\em Kepler} data covering Q1 through 5 only
(13 months), I will here present results based on Q1 through 13 (3 years and 1 month),
increasing the frequency resolution threefold and lowering the mean noise level
in the Fourier transform (FT) from 13.7 to 6.4 parts per million (ppm).

\begin{figure*}[!t]
\centering
\includegraphics[width=\textwidth]{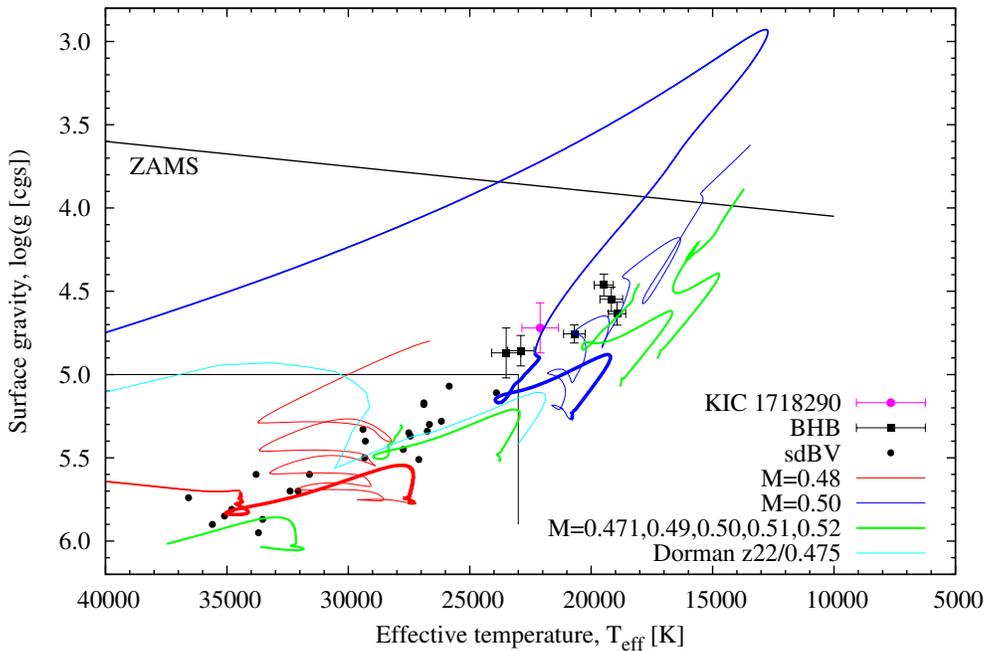}
\caption{
The \teff\--\logg\ plane with evolutionary models and some sample stars.
}
\label{fig:tgplot}
\end{figure*}

\section{Spectroscopy}

We reanalysed the original WHT spectrum of Largo presented in \citetalias{ostensen12b} using
the TLUSTY NLTE models of \citet{Nemeth13}, finding the same physical parameters
within the errors as the ones found in \citetalias{ostensen12b} using LTE models;
\teff\,=\,22520\,$\pm$\,150\,K, \logg\,=\,4.76\,$\pm$\,0.02\,dex, and
\logy\,=\-0.42\,$\pm$\,0.03. The errors of this fit are slightly lower as
as some He-lines have better defined profiles thereby reducing the formal fitting
errors. A further confirmation that the parameters of this unusual star is
correct is shown in the analysis of \citet{green13} in these proceedings.
Fig.~\ref{fig:tgplot} shows the location of Largo in the \teff/\logg\ plane.
The target and six other stars (PG\,0229+064, PG\,0848+186, PG\,1400+389, PG\,2356+167,
Balloon 82000001 and CPD--20$^\circ$1123)
are plotted with error-bars indicating three times the formal-fitting errors.
Five of these are also reanalysed in \citet{green13}, but the figure still shows
the old LTE values from \citetalias{ostensen12b} for consistency.

Inside the rectangular outline in Fig.~\ref{fig:tgplot}, the pulsating hot subdwarfs
from the \kep\ sample are
plotted with their physical parameters from \citet{ostensen10b,ostensen11b},
and further down the EHB we include some V368\,Hya
pulsators from \citet{sdbnot} that were fitted
on a similar LTE grid as the other stars.
Note that the number of stars marked in Fig.~\ref{fig:tgplot} is in no way representative
of the population densities of these stars. The EHB stars in the figure are just 24 stars
out of more than 2000 known sdBs \citep{ostensen06}, while the BHB stars indicated
include the majority of stars similar to Largo described in the literature.
Thus, as pointed out by \citet{saffer94,saffer97},
what is known as the second Newell gap from studies of globular clusters \citep{newell76}
is not a forbidden region on the BHB nor due to selection effects, but rather a very
underpopulated region among field stars. Some globular clusters are known to show
unusual blue horizontal branch morphologies, which may be a helium and/or metalicity
effect. The low galactic latitude of Largo may indicate that it belongs to a different
population than the majority of hot subdwarfs in the field.

The evolutionary tracks plotted in the figure were computed with MESA
\citepalias[see][for details]{ostensen12b}, all using the same core model with
different envelope thicknesses. The tracks demonstrate that Largo most likely
has a total mass close to 0.50\,\msol. Note the clear displacement of the
observed EHB+BHB stars compared to the core-Helium burning stage
of the evolutionary tracks where the stars spends most of their HB
lifetime (the knee-shaped part of the track plotted with thick curves).
It is not clear if this discrepancy originates with the spectroscopic
models, the fitting procedure, the evolutionary models, or a combination of
those. 


\begin{figure*}[t]
\centering
\includegraphics[width=\textwidth]{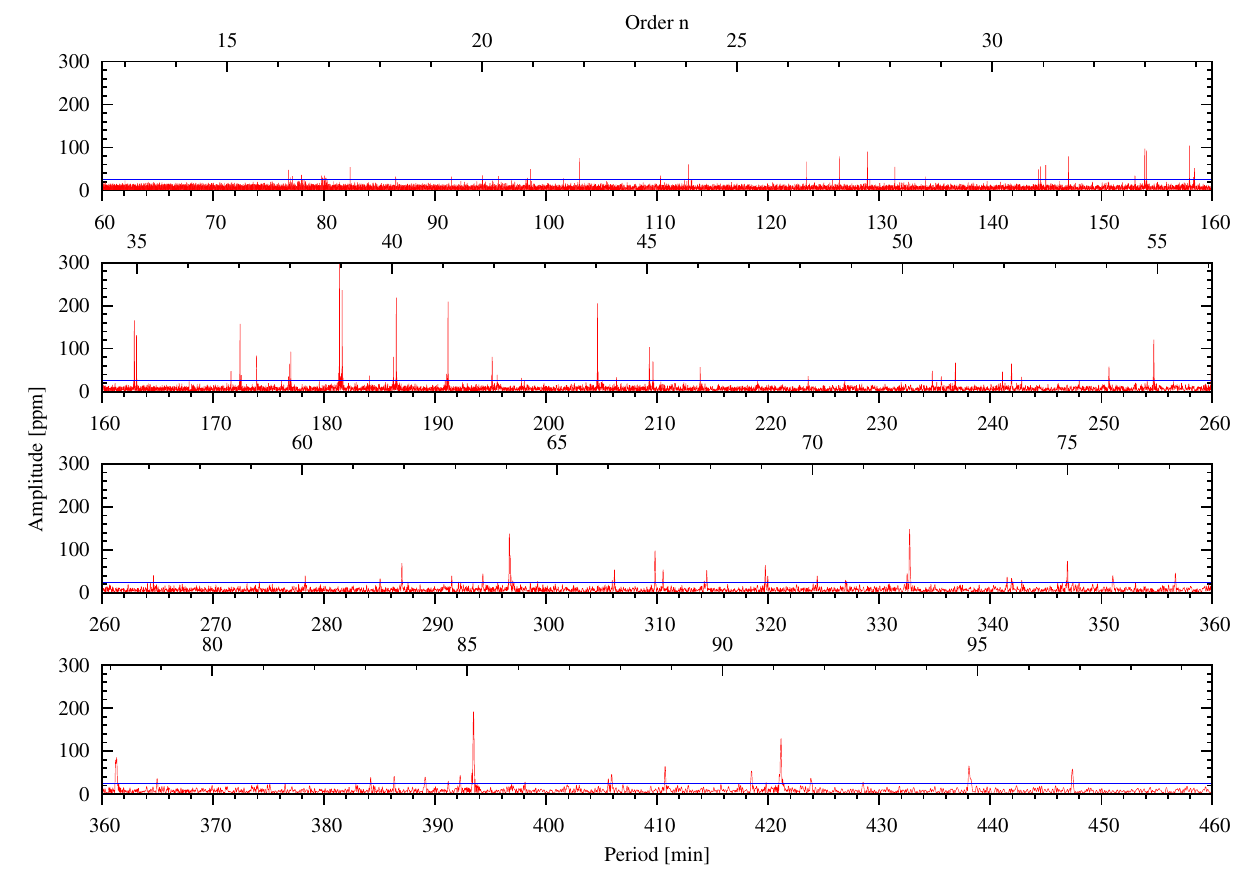}
\caption{
The FT of the \kep\ light curve of Largo, including
data from Q1 through Q13, plotted with period on the abscissa.
The 4-$\sigma$ level is indicated with a horizontal line.
On the upper axis the approximate radial order $n$ of the
\ellone\ modes is given,
according to the asymptotic relation $\Pi$\,=\,$n\cdot\Delta\Pi+\epsilon$
= $n\cdot275.7+140$\,s.
}
\label{fig:ftper}
\end{figure*}

\section{Kepler data}

\begin{figure*}[t]
\centering
\includegraphics[width=\textwidth]{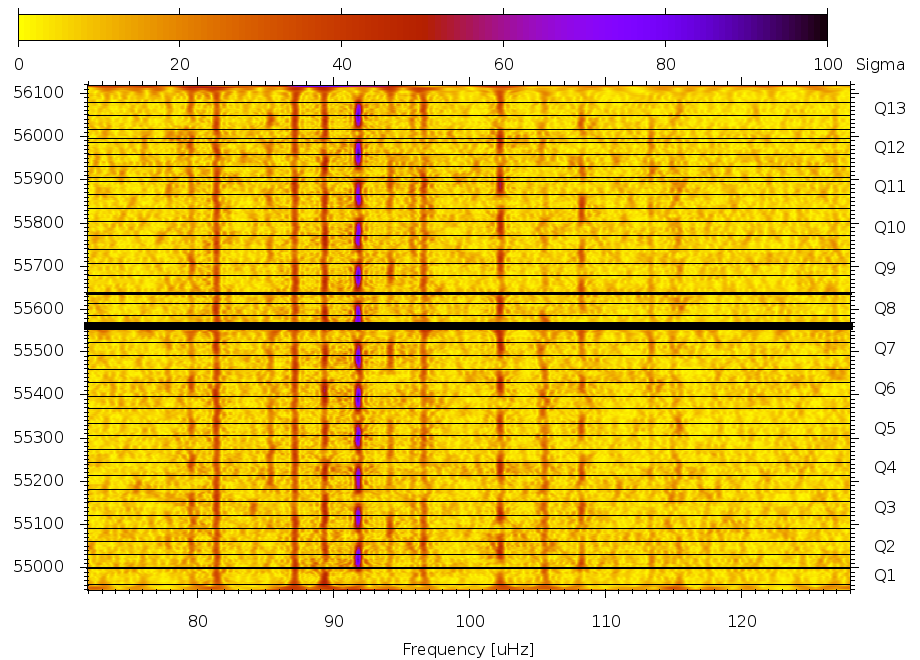}
\caption{
Running FT of a section around the highest peak in the frequency
spectrum ($f_1$\,=\,92\,\uHz, corresponding to 181\,s in the periodogram).
}
\label{fig:running}
\end{figure*}

We processed the additional eight quarters of \kep\ data in the same way
as described in \citetalias{ostensen12b}, and proceeded to identify 119 frequencies
with amplitudes higher than 30\,ppm, more than double the number of peaks
identified in the discovery paper. The large number of modes that follow
a regular period spacing can be easily discerned in Fig.\,\ref{fig:ftper}.
It is hard to discern in the periodogram, but most of the high-amplitude peaks
are clear doublets. If one produces a running FT as in Fig.\ref{fig:running}
(by chopping up the light curve into 30-day chunks, taking the FT of each
chunk, and stacking them along the ordinate axis),
the now unresolved doublets emerge with a distinctive beat pattern.

\begin{figure*}[t]
\centering
\includegraphics[width=\textwidth]{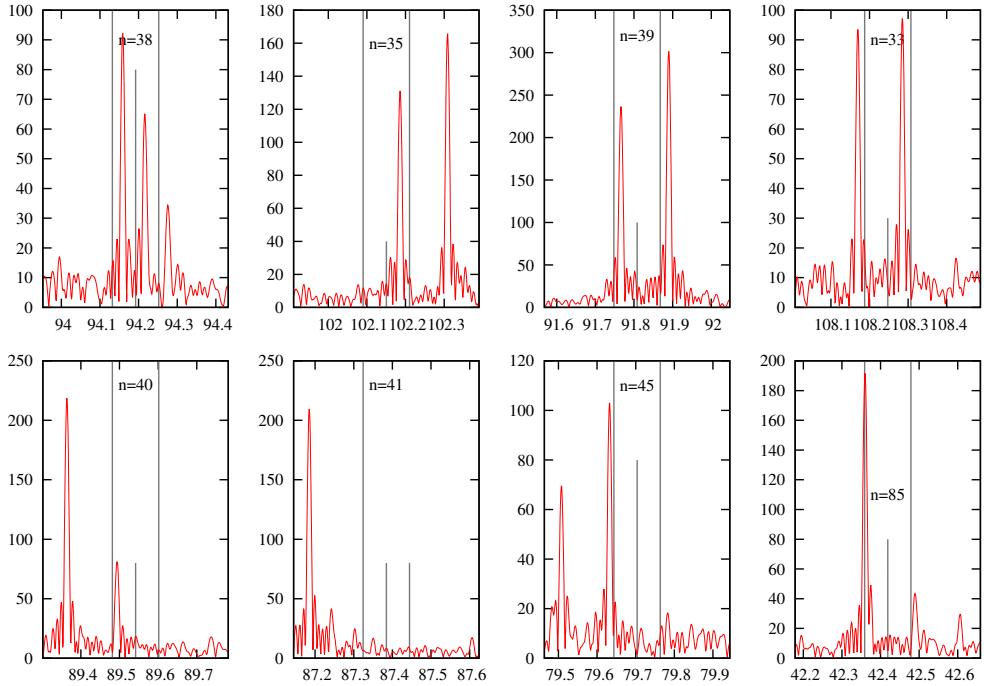}
\caption{
Zoom-in views of some of the strongest peaks in Fig.\ref{fig:ftper}.
The gray vertical bars show the locations of the \ellone\ triplets
as predicted by the asymptotic relation.
The splitting of the $m=\pm 1$ is 0.12\,uHz, which corresponds
to a rotation period of 96.5\,d.
Only one mode, $n$\,=\,38, shows a clear triplet. All the other
doublets are missing the central \emmzero\ component.
}
\label{fig:ftzoom}
\end{figure*}

In Fig.~\ref{fig:ftzoom} we show zoom-ins views of some of the strongest
peaks in the FT, the majority of which appears as doublets.
The splitting of these is 0.12\,\uHz, which corresponds to 96 days
as is the period of the beat pattern seen in Fig.~\ref{fig:running}.

\begin{figure}[t!]
\centering
\includegraphics[width=\textwidth]{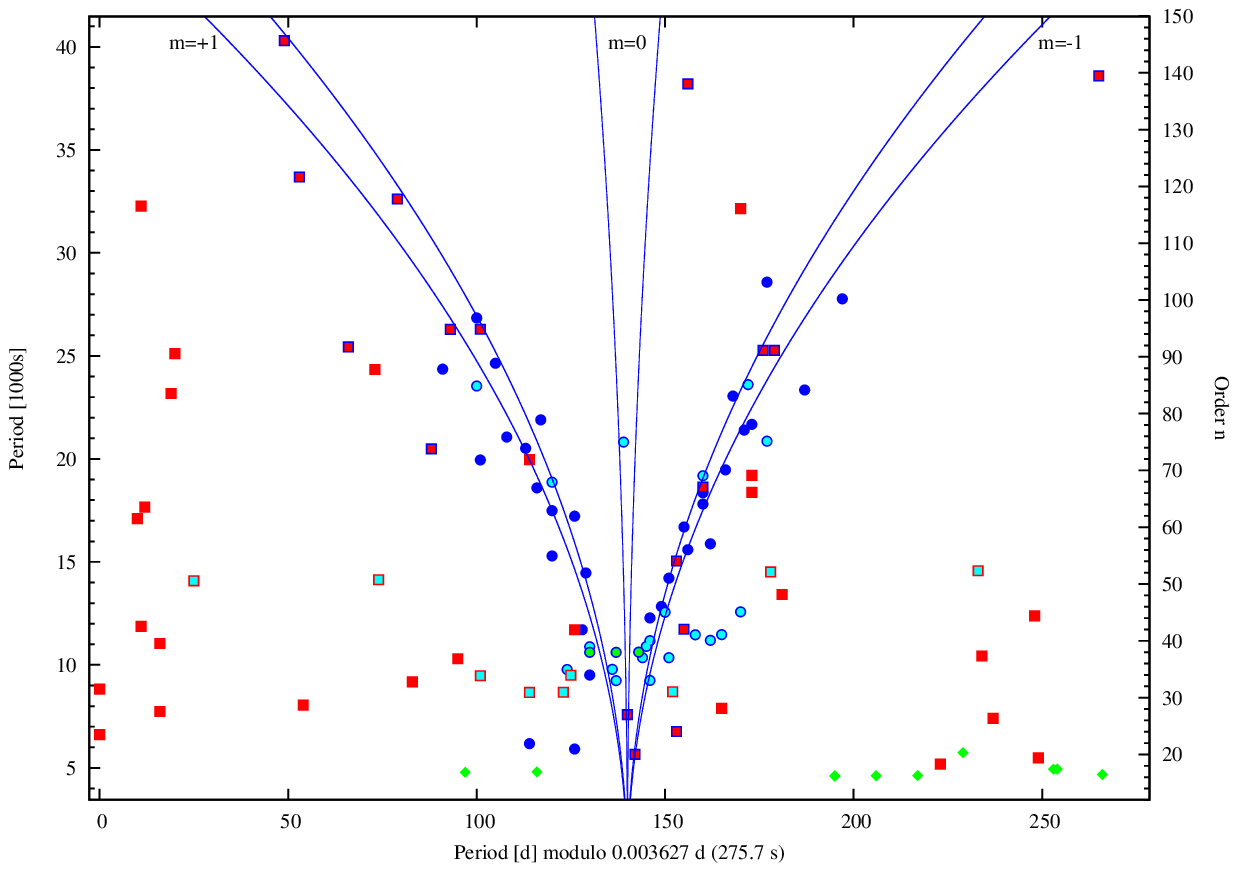}
\caption{
\'{E}chelle diagram for \ellone. The peaks are plotted modulo 275.7\,s.
Periods which are part of doublets are marked with round cores. The bullets indicate
\ellone\ modes and boxes \elltwo\ modes. Periods that may belong
to either sequence are marked with hollow boxes.
A cluster of points below $\sim$5000\,s (diamonds) do not fit any sequence.
The three curve pairs indicate the width of an \ellone\ triplet splitting,
with the separation of each pair indicating the formal frequency resolution
for the 3-year dataset.
}
\label{fig:echelle}
\end{figure}

\section{The fabulous \'Echelle diagram}

While the periodogram and running FT are useful tools for easily
revealing regular periodicities in frequency spacings, the
favourite tool for asteroseismologists have long been the \'echelle
diagram. The classical \'echelle diagram is constructed for $p$-mode
pulsators by determining the `large spacing'´, $\Delta\nu,$ and plotting all detected
modes with the frequency modulo $\Delta\nu$ on the abscissa and frequency
on the ordinate axis. In these \'echelle diagrams, modes of equal $\ell$ will
make roughly vertical stacks in the asymptotic limit. For $g$-modes in the
asymptotic limit modes are roughly evenly spaced in period, with
a spacing, $\Delta\Pi$, that is different for each $\ell$. The sequences follow
$\Pi_{\ell,n}$\,=\,$n\cdot\Pi_{0}/\sqrt{\ell(\ell+1)}+\epsilon$, so spacing for
the sequence of \elltwo\ is $1/\sqrt{3}$ times that of \ellone, and so on.
Thus, unlike the $p$-mode \'echelle diagram where modes of all $\ell$
stack with the same spacing, one must make one diagram for each $\ell$ with
different folding periods.

Fig.~\ref{fig:echelle} shows the \'echelle diagram for \ellone\ constructed
using $\Delta\Pi_{\ell=1}$\,=\,275.7\,s. Note that this is 0.5\,s less than
the 276.3\,s period used in \citetalias{ostensen12b}, which changes the result
quite significantly. The $\Delta\Pi_{\ell=1}$ inferred in the discovery paper
was achieved by attempting to get as many modes as possible to line up on
the \ellone, \emmzero\ sequence. Fig.~\ref{fig:echelle} shows why this produced
the wrong result. In the new interpretation, the majority of modes fall on the
\emmone\ curves with hardly any \emmzero\ modes present.
The standard explanation why the \emmzero\ modes are mostly absent would be that
the viewing angle is unfavourable, {\em i.e.} we are observing the star with
the pulsation axis nearly perpendicular to the line of sight. However, a
close inspection of the strongest \ellone\ modes in Fig.~\ref{fig:ftzoom}
does not readily support this interpretation. The upper left panel ($n=38$) is
a very clear triplet, and the central component is far from being weak. All the
other panels, except perhaps $n=45$ show no trace of the central component,
implying a suppression of at least 90\%. Furthermore, the majority of the
g-mode pulsating sdBs in the Kepler fields show far more doublets than triplets
which should not be the case if was caused by the random viewing angle.
Further deepening this mystery is the fact that the doublets are almost entirely
clustered in the $n=35$ to 45 region. From $n=45$ and beyond $n=100$ only 3
of about 40 \ellone\ modes are possible doublets, the rest are single and
evenly distributed between the $m=+1$ and $m=-1$ bands.

\begin{figure}[t!]
\centering
\includegraphics[width=\textwidth]{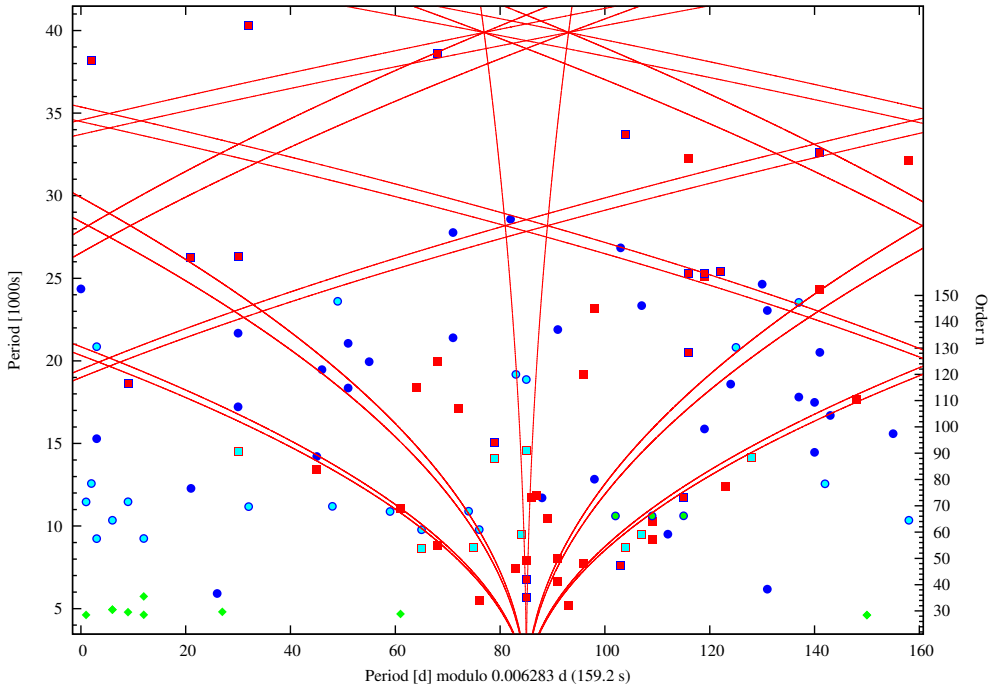}
\caption{
\'{E}chelle diagram for \elltwo\ (modulo 159.2\,s).
Symbols and curves are as in Fig.~\ref{fig:echelle}.
Note that while the quintuplet splittings diverge much earlier
than for \ellone\ (due to the smaller folding interval), it is
still clear that a lot of modes in the region between 5000 and
15\,000\,s that clearly does not fit the \ellone\ sequence, fits
well on the \elltwo\ curves, and mostly on the outer \emmtwo\ ones.
At periods above 15\,000\,s the \'echelle diagram becomes less useful
as a mode identification tool as anything can fit within 3$\sigma$ of
a curve.
}
\label{fig:echelle2}
\end{figure}

Proceeding to the \elltwo\ modes (Fig.~\ref{fig:echelle2}) the picture is less
clear, but some of the same features appears to be present. There are a few
likely multiplets mostly in the same region around 10\,000\,s where the \ellone\
multiplets were found, but the majority of \elltwo\ modes show only a single excited
component.
Again, the outermost tracks ($m=\pm 2$) appears to be favored, which would be the
case when the pulsation axis is viewed equator-on. 
The last plot, Fig.~\ref{fig:echelle2}, shows the \'echelle diagram for
\ellthree. It is clear that no obvious clustering appears on these tracks, but 
the curves quickly becomes so dense that no conclusions can be made.

\begin{figure}[t!]
\centering
\includegraphics[width=\hsize]{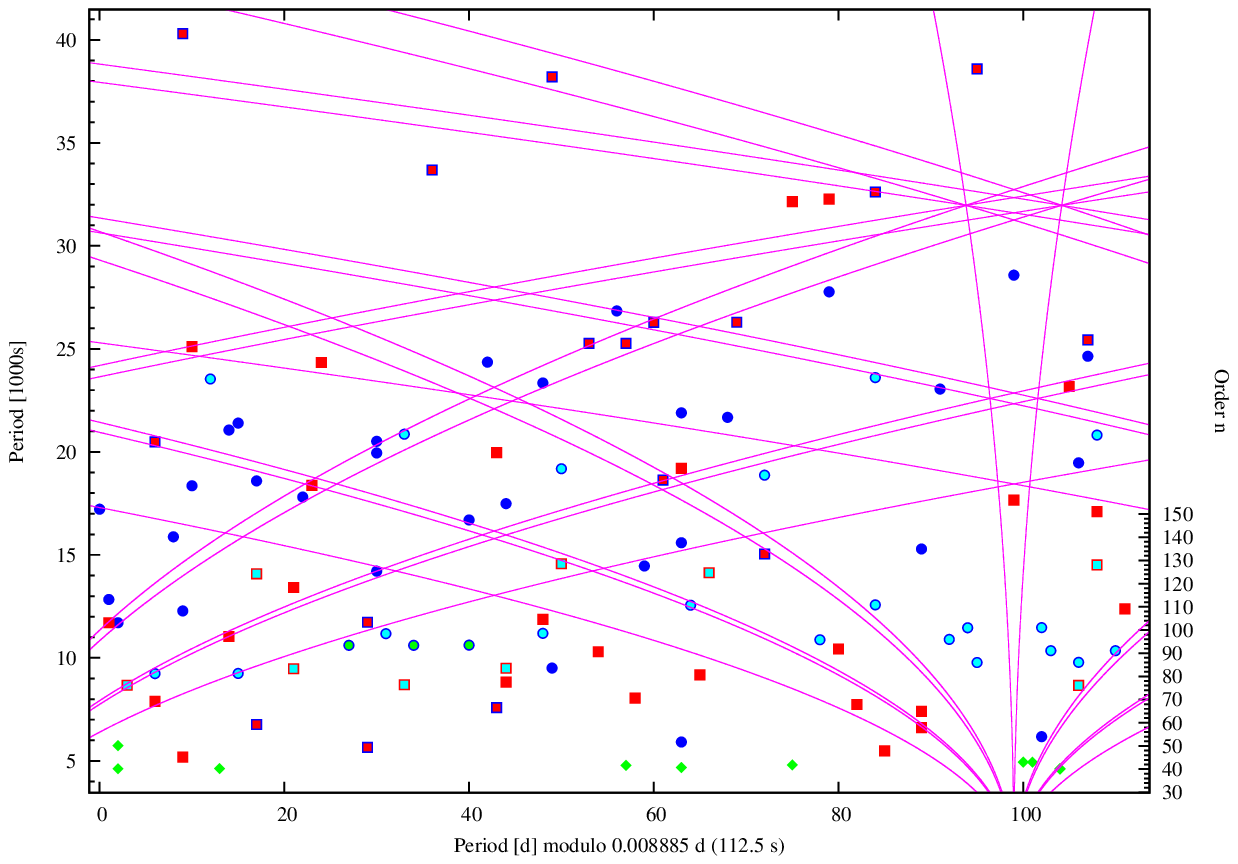}
\caption{
\'{E}chelle diagram for \ellthree\ (modulo 112.5\,s), again with
symbols are as in Fig.~\ref{fig:echelle}.
There is no convincing evidence for any clustering on the \ellthree\
curves. The points that fits neither \ellone\ nor 2 (diamonds below 5000\,s)
are clearly far too densely spaced to be consistent with \ellthree.
Note that the \'echelle diagram is cannot be used to exclude \ellthree\
identification for any mode above $\sim$10\,000\,s. Since $\epsilon$ is not
determined for \ellthree, even some of the points that are located inside
the excluded zone in the bottom left part of the diagram can still be 
\ellthree\ modes.
}
\label{fig:echelle3}
\end{figure}

\section{Conclusions}

Largo is an exceptionally interesting target, providing an asteroseismic
connection between the blue horizontal branch stars at \logg\,$<$\,5 and
the V1093\,Her stars on the EHB. Being cooler and larger than the EHB
pulsators, the excited modes of Largo spans periods between about 5000\,s
and 40\,000\,s, a range perfectly suited for \kep\ long-cadence observations.
The fabulous persistence of \kep\ monitoring provides a wealth of information
that invites the application
of more sophisticated tools than traditional Fourier analysis.
The running FT reveals distinctive beat patterns that readily identifies
multiplets and allows beating modes to be distinguished from random amplitude
variability. 
Furthermore, the g-mode \'echelle diagram provides an excellent tool for
studying \ellone\ and \elltwo\ modes in such stars, thanks to its slow rotation
rate of $\sim$100 days.
For \ellthree\ and above the \'echelle diagram is
not really useful as the folding period becomes too short even at the slow
rotation rate found for Largo.

\acknowledgements
The research leading to these results has received funding from the European
Research Council under the European Community's Seventh Framework Programme
(FP7/2007--2013)/ERC grant agreement N$^{\underline{\mathrm o}}$\,227224
({\sc prosperity}), as well as from the Research Council of K.U.Leuven grant
agreement GOA/2008/04.

\bibliographystyle{asp2010}
\bibliography{sdbrefs}

\end{document}